\documentclass[pra,twocolumn,floatfix,nofootinbib]{revtex4}

\usepackage{graphicx}
\usepackage{amsmath}
\usepackage{amsfonts}
\usepackage{amssymb}
\usepackage{dcolumn}
\usepackage{bm}

\newcommand{\be}{\begin{equation}}
\newcommand{\ee}{\end{equation}}
\newcommand{\ba}{\begin{eqnarray}}
\newcommand{\ea}{\end{eqnarray}}

\newenvironment{proof}[1][Proof]{\noindent\textbf{#1.} }{\ \rule{0.5em}{0.5em}}
\newtheorem{myproblem}{Problem}

\newtheorem{myproposition}{Proposition}

\newtheorem{mydefinition}{Definition}
\newtheorem{mytheorem}{Theorem}

\newtheorem{mycorollary}{Corollary}

\begin{document}

\title{On the Quantum Computational Complexity of the Ising Spin Glass
  Partition Function and of Knot Invariants}
\author{Daniel A. Lidar}
\affiliation{Chemical Physics Theory Group, Chemistry Department, and Center for Quantum
Information and Quantum Control, University of Toronto, 80 St. George St.,
Toronto, Ontario M5S 3H6, Canada}

\begin{abstract}
It is shown that the canonical problem of classical statistical
thermodynamics, the computation of the partition function, is in the case of 
$\pm J$ Ising spin glasses a particular instance of certain simple sums
known as quadratically signed weight enumerators (QWGTs). On the other hand
it is known that quantum computing is polynomially equivalent to classical
probabilistic computing with an oracle for estimating QWGTs. This suggests a
connection between the partition function estimation problem for spin
glasses and quantum computation. This connection extends to knots and graph
theory via the equivalence of the Kauffman polynomial and the partition
function for the Potts model.
\end{abstract}

\maketitle

\section{Introduction}

Feynman famously conjectured that unlike classical computers, quantum
computers should be efficient at simulating quantum mechanics
\cite{Feynman:QC,Feynman:QC1}. This was verified by Lloyd
\cite{Lloyd:96}, and further developed by several other authors, who
demonstrated exponential speedup over the best known classical
algorithms for a variety of quantum mechanical problems such as
solution of the constant potential Schr\"{o}dinger and Dirac equations
using lattice gas automata
\cite{Meyer:96,Meyer:97,Boghosian:97c,Boghosian:97a}, solution of the
Schr\"{o}dinger equation in the circuit model
\cite{Wiesner:96,Zalka,Zalka:98}, simulation of fermionic systems
\cite{Abrams:97,Ortiz:00}, computation of the thermal rate constant of
chemical reactions \cite{Lidar:98RC}, computation of correlation
functions \cite{Terhal:00}, simulation of topological field theories
\cite{Freedman:00}, and simulation of pairing Hamiltonians such as the
BCS model \cite{WuByrdLidar:01}. A naturally related question is
whether quantum computers can efficiently solve problems in
\emph{classical physics}. This question was first raised, and
partially answered, in the context of Ising spin glasses
\cite{Lidar:PRE97a}. It has recently received renewed attention in the
context of hydrodynamics \cite{Yepez:01,Meyer:02} (with polynomial
speedups), chaos \cite{Georgeot:01a,Georgeot:01b,Terraneo:03} (with
exponential speedups, though some of these have been challenged
\cite{Zalka:01}), and knot theory \cite{Kauffman:01} (so far without
speedup), which has a deep connection to classical statistical
mechanics \cite{Kauffman:book}.

Here we revisit classical Ising spin glasses and also address knot theory.
The canonical problem of classical statistical thermodynamics is the
calculation of the partition function $Z$. For a system in thermodynamic
equilibrium, if the partition function is known, one can obtain
exact results for all thermodynamic quantities such as the
magnetization, susceptibility and specific heat. Models for which analytical
calculations of this type can be performed include a variety of one
dimensional (1D) models and the 2D Ising model \cite{Onsager44,Baxter:book}.
However, for most systems of interest, including the 3D Ising model and most
Ising spin glass models, no analytical calculation of the partition function
is available \cite{Green}. We consider classical spin systems, in particular
the Ising model \cite{Ising} in which each spin has two states $\sigma_{i}=\pm 1$%
, and spins interact pairwise with an interaction energy of the form $%
J_{ij}\sigma_{i}\sigma_{j}$. From a computational complexity perspective this provides
a rich class of problems. In particular, the problem of finding the ground
state of the short range 3D Ising spin glass (quenched random $J_{ij}$), as
well as the fully antiferromagnetic (all $J_{ij}=-|J|$) 2D Ising model in
the presence of a constant magnetic field was shown by Barahona to be
NP-hard, by a mapping to problems in graph theory.\footnote{NP-hard problems are those whose
proposed solution cannot even be \emph{verified} using a nondeterministic
Turing machine in polynomial time; e.g., is the
proposed ground state truly the ground state?} On the other hand, it is known that there exists
a fully polynomial randomized approximation scheme for the ferromagnetic
Ising model (all $J_{ij}=|J|$), on arbitrary graphs \cite{Jerrum:90}.

The problem of sampling from the Gibbs distribution of the $J_{ij}=\pm J$
(with random signs) spin glass on a quantum computer was addressed in \cite{Lidar:PRE97a}. A linear-time algorithm was found for the construction of
the Gibbs distribution of configurations in the Ising model, including \emph{partially} frustrated models. A magnetic field can be incorporated as well
without increase in the run-time. The algorithm was designed so that each
run provides one configuration with a quantum probability equal to the
corresponding thermodynamic weight. I.e., the probabilities of measuring
states are ordered by the energies of the corresponding spin configurations,
with the ground state having the highest probability. Therefore the
partition function is approximated efficiently and statistical averages
needed for calculations of thermodynamic quantities obtained from the
partition function, are approximated in the fastest converging order in the
number of measurements. Unlike Monte Carlo simulations on a classical
computer, consecutive measurements on a quantum computer can be totally
uncorrelated. Thus the algorithm neither suffers from critical slowing down
(a polynomial slowdown in Monte Carlo moves associated with large correlated
spin clusters forming near the critical temperature) \cite{Swendsen87}, nor
gets stuck in local minima. This \emph{uniform} performance is an advantage
over the best known classical algorithms, which are tailored to specific
lattices or graphs \cite{Swendsen87}. However, the main problem of the
algorithm is the limited control it offers in the construction of a \emph{specific} realization of bonds on the Ising lattice. Indeed, since the
run-time of the algorithm is linear, it is reasonable to suspect that it
cannot simulate a hard instance of an Ising spin glass.

A completely different approach to sampling from the Gibbs distribution for
the \emph{ferromagnetic} Ising model was recently developed by Nayak,
Schulman and Vazirani (NSV) \cite{Nayak:unp}, which however does not appear
to provide a speedup over the best known classical algorithm \cite{Randall:99}. The VNS algorithm for the ferromagnetic Ising model uses an
interesting representation of the partition function as a discrete Fourier transform,
in conjunction with a Markov chain sampling procedure. Two of the main
results of the present paper are (i) the generalization of this Fourier
transform representation to the $\pm J$ spin glass case, (ii) (the central
result) a connection of this representation to certain
simple sums known as \textquotedblleft quadratically signed weight
enumerators\textquotedblright\ (QWGTs). Let us now motivate the importance
of result (ii).

In virtually all previous work on simulation of physical systems on quantum
computers \cite{Lloyd:96,Meyer:96,Meyer:97,Boghosian:97c,Boghosian:97a,Wiesner:96,Zalka,Zalka:98,Abrams:97,Ortiz:00,Lidar:98RC,Terhal:00,Freedman:00,WuByrdLidar:01,Lidar:PRE97a,Yepez:01,Meyer:02,Georgeot:01a,Georgeot:01b,Terraneo:03,Zalka:01,Kauffman:01}%
, the approach pursued was one of attempting to find a \emph{concrete
algorithm} for a specific simulation problem. A fruitful alternative is to
consider instead the question of the \emph{complexity class} that the
simulation problem belongs to. We do this here by following a lead due to
the Knill and Laflamme (KL): KL showed that quantum computing is
polynomially equivalent to classical probabilistic computing with an oracle
for estimating QWGTs \cite{Knill:01a}. Combined with our results (i),(ii),
this suggests that the quantum computational complexity of sampling from the
Gibbs distribution for the Ising spin glass problem can be understood in
terms of QWGTs. However, we have unfortunately not yet been able to
establish the connection at this level. Nevertheless, the \emph{possibility}
of a (spin glass)-(QWGTs)-(quantum computation) connection is sufficiently
tantalizing to point it out in detail. We hope that by expressing the
partition function as a QWGT we have taken the first step in a direction
that will allow future research to explore the important question of the
quantum computational complexity of the Ising spin glass problem.

In fact, the connections do not end here. There is a rich inter-relation
between classical statistical mechanics and topology, in particular the
theory of classification of knots. The first such connection was established
by Jones \cite{Jones:89}, who discovered knot invariants (the Jones'
polynomial) during his investigation of topological properties of braids 
\cite{Jones:85}. It is known that classical evaluation of the Jones'
polynomial is \#P-hard \cite{Jaeger:90}. The connection between knots and
models of classical statistical mechanics was greatly embellished by
Kauffman \cite{Kauffman:book}. Here we will exploit this connection to show
[our main result (iii)] that the evaluation of another knot invariant, the
Kauffman polynomial, can also be cast in some cases as a QWGT evaluation
problem. Thus a quantum algorithm for QWGT evaluation should shed light on
the quantum computational complexity of knot invariants, a subject which has
been explored by Freedman et al. \cite{Freedman:01} and by Kauffman and
Lomonaco \cite{Lomonaco}. Knot invariants are, in turn, also tightly related
to graph theory; e.g., the graph coloring problem can be considered an
instance of evaluation of the Kauffman polynomial, via the Tutte polynomial 
\cite{Kauffman:book}.

Mathematically, the reason that these seemingly unrelated subjects are
all inter-related is due to the fact that key properties can be expressed,
in all cases, in terms of certain polynomials. While these polynomials
originate from widely distinct problems, from the point of view of
computational complexity their evaluation is one and the same problem, much
in the same spirit as the fact that solving one problem in the class of
NP-complete problems solves them all \cite{Papadimitriou:book}. The present
work contributes to this unification.

The structure of this paper is as follows. In Sec.~\ref{Fourier} we derive our
first main result: we rewrite the Ising spin glass partition function for
an arbitrary graph as a discrete Fourier transform. This motivates the consideration
of the partition function evaluation problem in terms of its computational
complexity, which we formalize in Sec.~\ref{complexity}. We then review
QWGTs in Sec.~\ref{QWGT}. In Section~\ref{connection} we derive our second
main result: the connection between the evaluation of the Ising spin glass
partition function and QWGTs. In Sec.~\ref{knots} we continue the program of
connecting disparate objects to the problem of QWGT evaluation, and
obtain our third main result: we show
that the Kauffman polynomial too can be expressed a QWGT. We do this after
first reviewing the connection between knots and classical statistical
mechanics. Section~\ref{conclusion} concludes. Some further observations
concerning the representation of the partition function are collected in
Appendix~\ref{further}.

\section{Fourier Transform Representation of the Partition Function}
\label{Fourier}
Let $G=(E,V)$ be a finite, arbitrary undirected graph with $|E|$ edges and $%
|V|$ vertices. Identify each vertex $i\in V$ with a classical spin ($\sigma
_{i}=\pm 1$) and each edge $(i,j)\in E$ with a bond ($J_{ij}=\pm J$). Denote
a given spin configuration by $\sigma =(\sigma _{1},\sigma _{2},...,\sigma
_{|V|})$ and a bond configuration by $(J_{12},...,J_{ij},...)$. We assume
that the bond configuration is chosen at random and then remains fixed
(\textquotedblleft quenched randomness\textquotedblright ). The Hamiltonian
of the system is

\begin{equation}
H(\sigma)=-\sum_{(i,j)\in E}J_{ij}\sigma _{i}\sigma _{j}.  \label{eq:H}
\end{equation}
(We remark on the case with a magnetic field in Appendix \ref{appA}.) The
probability of the spin configuration $\sigma$ in thermal equilibrium at
temperature $T$ is given by the Gibbs distribution:

\begin{equation}
P(\sigma )={\frac{1}{Z}}W(\sigma ),  \label{eq:Gibbs}
\end{equation}%
where the Boltzmann weight is 
\begin{equation}
W(\sigma )=\exp [-\beta H(\sigma )],  \label{eq:Boltz}
\end{equation}%
$\beta =1/kT$ is the inverse temperature, and $Z$ is the partition function: 
\begin{equation}
Z(\{J_{ij}\})=\sum_{\sigma }\exp [-\beta H(\sigma )].  \label{eq:Z}
\end{equation}%
Now note the identity

\begin{equation}
e^{x}=\cosh (x)[1+\tanh (x)]
\end{equation}
and use it to rewrite the Boltzmann weight~(\ref{eq:Boltz}) as

\begin{equation}
W(\sigma)=\prod_{(i,j)\in E}\cosh (\beta J_{ij}\sigma _{i}\sigma
_{j})[1+\tanh (\beta J_{ij}\sigma _{i}\sigma _{j})].
\end{equation}
Let

\begin{equation}
J_{ij}=q_{ij}J,
\end{equation}
where $q_{ij}=\pm 1$ is a quenched random variable. Since $\cosh (x)=\cosh
(-x)$ and $\tanh (x)=-\tanh (-x)$ we find

\begin{equation}
W(\sigma)=\Theta \prod_{(i,j)\in E}(1+q_{ij}\sigma _{i}\sigma _{j}\lambda ),
\end{equation}
where

\begin{equation}
\Theta ={[\cosh (\beta J)]^{|E|},}  \label{eq:Theta}
\end{equation}%
and 
\begin{equation}
\lambda =\tanh (\beta J).  \label{eq:lambda}
\end{equation}%
Next expand out the product to obtain

\begin{eqnarray}
P(\sigma ) &=&\Theta \lbrack 1+\lambda \sum_{(i,j)\in E}\sigma
_{i}q_{ij}\sigma _{j}  \notag \\
&&+\lambda ^{2}\sum_{(i,j),(k,l)\in E}(\sigma _{i}q_{ij}\sigma _{j})(\sigma
_{k}q_{kl}\sigma _{l})+....]  \label{eq:1}
\end{eqnarray}%
Note that $\lambda ^{k}$ is the coefficient in front of a sum containing $k$
bonds $q_{ij}$, which are not necessarily all connected. For example, the
term $q_{ij}\sigma _{i}\sigma _{j}q_{kl}\sigma _{k}\sigma _{l}$ where all
indexes differ represents two unconnected bonds (Fig.~\ref{fig:graphs}a),
whereas if $j=k$ the same term represents two connected bonds sharing one
spin (Fig.~\ref{fig:graphs}b). Let $b\subseteq E$ denote such a subgraph,
with $k=|b|$ edges. There are $\binom{|E|}{|b|}$ ways of choosing subgraphs
with $|b|$ edges. Thus the total number of subgraphs is:

\begin{figure}[tbp]
\hspace{10cm} \includegraphics[height=10cm,angle=270]{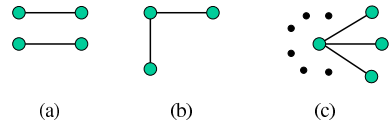} \vspace{%
-6cm}
\caption{Illustration of subgraphs.}
\label{fig:graphs}
\end{figure}

\begin{equation}
\sum_{|b|=0}^{|E|}\binom{|E|}{|b|}=2^{|E|}.
\end{equation}%
This suggests to label the subgraphs in a binary fashion (to be explained
below): Let $b=(b_{1},b_{2},...,b_{|E|})$ (where $b_{j}\in \{0,1\}$) be the
binary number of subgraph $b$. The numbering is such that $b_{j}=1$ ($0$)
indicates the presence (absence) of edge number $j$ of $G$. Further, we will
use the convention that first all $|E|$ single-edge subgraphs are counted
(i.e., vectors $b$ with a single $1$ entry, all the rest $0$), then all $%
\binom{|E|}{2}$ double-edge subgraphs (vectors $b$ with two $1$ entries, all
the rest $0$), etc. Thus, e.g., $b=(0,0,...,0,1)$ could corresponds to the
subgraph containing only the first edge (or bond): $J_{12}$, whereas $%
b=(0,0,...0,1,0,1)$ could correspond to the subgraph containing only $J_{12}$
and $J_{34}$. Note that $|b|$ is the Hamming weight of $b$. Since the total
number of subgraphs is $2^{|E|}$, the above numbering scheme is a one-to-one
covering of the subgraph space, and moreover, the subgraphs are labeled in
increasing order of $|b|$.

Next note that in Eq.~(\ref{eq:1}), a spin $\sigma _{i}$ may appear more
than once in sums of order $k\geq 2$, in fact as many times as the number of
bonds emanating from $\sigma _{i}$, which we denote by $\mathrm{deg}_{b}(i)$%
. Then $\sigma _{i}^{\mathrm{deg}_{b}(i)}$ is the contribution of spin $i$
in subgraph $b$ to the product in the sum of order $k=|b|$ in Eq.~(\ref{eq:1}%
). Collecting all the observations above, it follows that Eq.~(\ref{eq:1})
can be rewritten as:

\begin{equation}
P(\sigma )=\Theta \sum_{b}\lambda ^{|b|}\prod_{i\in V(b)}\sigma _{i}^{%
\mathrm{deg}_{b}(i)}\prod_{(i,j)\in b}q_{ij}.  \label{eq:2}
\end{equation}%
Here $V(b)$ denotes the set of vertices in subgraph $b$.

Next introduce the \emph{parity} vector ${\alpha }^{b}=(\alpha
_{1}^{b},\alpha _{2}^{b},...,\alpha _{|V|}^{b})$ of a subgraph, with
components $\alpha _{i}^{b}=1$ if ($i\in V(b)$ and $\mathrm{deg}_{b}(i)$ is
odd), $\alpha _{i}^{b}=0$ otherwise. Then clearly $\sigma _{i}^{\mathrm{deg}%
_{b}(i)}=\sigma _{i}^{\alpha _{i}^{b}}$. At this point it is more convenient
to transform to a binary representation for the spins as well. Let:

\begin{equation}
s_{i}={\frac{1-\sigma _{i}}{2}}
\end{equation}
be the components of the binary vector of spin values $%
s=(s_{1},s_{2},...,s_{|V|})$. We have $\sigma _{i}=(-1)^{s_{i}}$ so that $%
\sigma _{i}^{\mathrm{deg}_{b}(i)}=(-1)^{\alpha _{i}^{b}s_{i}}$. Therefore:

\begin{equation}
  \prod_{i\in V(b)}\sigma _{i}^{\mathrm{deg}_{b}(i)}=(-1)^{{\alpha}^{b}\cdot s},
  \label{eq:3}
\end{equation}
where \textquotedblleft $\cdot $\textquotedblright\ stands for the (mod 2)
bit-wise scalar product. The same change can be
affected for the bonds by introducing:

\begin{equation}
w_{ij}={\frac{1-q_{ij}}{2}},
\end{equation}
so that the binary vector $w = (w_{12},w_{13},...)$ of length $|E|$
specifies whether edge $(i,j)$ supports a ferromagnetic ($w_{ij} = 0$) or
antiferromagnetic ($w_{ij} = 1$) bond. Again, $q_{ij} = (-1)^{w_{ij}}$, so
that:

\begin{equation}
  \prod_{(i,j)\in b}q_{ij}=(-1)^{b\cdot w}.
  \label{eq:4}
\end{equation}
Using Eqs.~(\ref{eq:3}),(\ref{eq:4}) in Eq.~(\ref{eq:2}) it is now possible
to rewrite the Boltzmann weight of a particular spin configuration $s$ as:

\begin{equation}
W(s) = \Theta \sum_{b}\lambda ^{|b|}(-1)^{{\alpha }^{b}\cdot
  s+b\cdot w}.
\label{eq:5}
\end{equation}
Now, the partition function is just the sum over all spin configurations.
Using Eq.~(\ref{eq:5}) we thus find:

\begin{myproposition}
\emph{The spin-glass partition function is a double discrete Fourier (or 
Walsh-Hadamard) transform, over the spin and subgraph variables}:
\begin{equation}
Z(w) = \Theta \sum_{s}\sum_{b}\lambda ^{|b|}(-1)^{{\alpha }^{b}\cdot
  s+b\cdot w}.
\label{eq:6}
\end{equation}
\end{myproposition}

Note that the sum over $s$ extends over $2^{|V|}$ terms, while the sum
over $b$ extends over $2^{|E|}$ terms. By changing the order of
summation, the sum over $s$ can actually be carried out, since:

\begin{equation}
\sum_{s}(-1)^{{\alpha }^{b}\cdot s}=2^{|V|}\delta _{({\alpha }^{b},0)},
\label{eq:7}
\end{equation}
where $\delta $ is the Kronecker symbol. A systematic procedure for finding
the parity vectors from the subgraphs vectors uses the \emph{incidence
  matrix} $A$. For any graph $G=(E,V)$ this matrix is defined as
follows ($v\in V$) \cite{Wilson:book}:

\begin{equation}
A_{v,(i,j)}=\left\{ 
\begin{array}{ll}
1 & \mbox{$(v=i \:\: {\rm and}\:\: (i,j)\in E)$} \\ 
0 & \mbox{${\rm else}$}%
\end{array}%
,\right.  \label{eq:A}
\end{equation}%
so $A$ is a $|V|\times |E|$ matrix of $0$'s and $1$'s. It is well known that
given $G$ and a specific subgraph $b$ \cite{Wilson:book},

\begin{equation}
  {\alpha }^{b}=A\cdot b.
  \label{eq:alpha}
\end{equation}
Combining Eqs.~(\ref{eq:Theta}),(\ref{eq:lambda}),(\ref{eq:6}),(\ref{eq:7}),(\ref{eq:alpha}), we
finally have:

\begin{equation}
Z(w)={\frac{2^{|V|}}{(1-\lambda ^{2})^{|E|/2}}}\sum_{b:\,A\cdot b=0}\lambda
^{|b|}(-1)^{b\cdot w}.  \label{eq:8}
\end{equation}%
In words, the sum over the subgraphs includes only those with zero overall
parity, i.e., those having an even number of bonds emanating from \emph{all}
spins. This immediately implies that \textquotedblleft
dangling-bond\textquotedblright\ subgraphs are not included in the sum. We
note that $Z(w)$ can also be rewritten as a power series in $\lambda $,
which is useful for a high-temperature expansion; this is discussed in App.~\ref{appB}. The representation~(\ref{eq:8}) allows us to establish a direct
connection with QWGTs, which is the subject of the quantum computational
complexity of $Z$, to which we turn next.

\section{Formulating the Computational Complexity of the Ising Spin Glass
Partition Function}

\label{complexity}

The most natural computational complexity class for quantum computation is
BQP: the class of decision problems solvable in polynomial time using
quantum resources (a quantum
Turing machine, or, equivalently, a uniform family of polynomial-size
quantum circuits) with bounder probability of error \cite{Aharonov:98,Aaronson-zoopage}. The class BQPP is the natural generalization
of BQP to promise problems \cite{Knill:01a}. Relative to the polynomial
hierarchy of classical computation, it is known that
BPP$\subseteq$BQP$\subseteq$PP$\subseteq$PSPACE, but none of these
inclusions is known to beproper \cite{Adleman:97}.

In order to address the quantum computational complexity of the spin glass
partition function we define:

\begin{mydefinition}
An \emph{instance} of the Ising spin glass problem is the data $\Delta \equiv
(G,w,\lambda )$.
\end{mydefinition}

Now, let $Z^{\Delta }(s)$ be the partition function for given data $\Delta $
and given spin configuration $s\in \{0,1\}^{|V|}$, and let
\begin{equation}
  Z_{k}^{\Delta}(s):\{0,1\}^{N}\rightarrow \{0,1\}
\end{equation}
be the $k^{\mathrm{th}}$ digit of $%
Z^{\Delta }(s)$. Let $\Gamma ^{\Delta }(s,k)$ be a quantum circuit that
takes as input the spin configuration $s$ and the digit location $k$, for
fixed data $\Delta $. Let $W_{\Gamma }=\prod_{i}U_{i}$ be an implementation
of $\Gamma ^{\Delta }$ in terms of some unitary operators $U_{i}$. Let the
circuit be designed so that the answer $Z_{k}^{\Delta }(s)$ is encoded into
the state of the first qubit, and let $\Pi _{1}=|0\rangle _{1}\langle 0|$ be
the projection onto state $|0\rangle $ of this qubit. Then the probability
of measuring the state $|0\rangle _{1}$ after the circuit was executed,
starting from the \textquotedblleft blank\textquotedblright\ initial state $|%
\mathbf{0}\rangle =|0\cdots 0\rangle $, is $\Pr [\Gamma ^{\Delta
}(s,k)]=\langle \mathbf{0}|W_{\Gamma }^{\dagger }\Pi _{1}W_{\Gamma }|\mathbf{%
0}\rangle $. We can now define:

\begin{mydefinition}
$Z_{k}^{\Delta }(s)\in $ BQP if there exists a classical polynomial-time
algorithm for specifying $\Gamma ^{\Delta }$ such that%
\begin{eqnarray*}
\Pr [\Gamma ^{\Delta }(s,k)] &\geq& \frac{2}{3}\hspace{3mm}\mathrm{if}\hspace{%
  3mm}Z_{k}^{\Delta }(s)=0\hspace{1cm}\mathrm{and}\\
\Pr [\Gamma
^{\Delta }(s,k)] &\leq& \frac{1}{3}\hspace{3mm}\mathrm{if}\hspace{3mm}%
Z_{k}^{\Delta }(s)=1.
\end{eqnarray*}
\end{mydefinition}

We can then formulate the following open problem:

\begin{myproblem}
\label{prob1}For which instances $\Delta $ of the Ising spin glass problem is evaluating
the partition function in BQP?
\end{myproblem}

A particularly promising way to attack this problem appears to be the
connection to QWGTs,  which we address next.

\section{Quadratically Signed Weight Enumerators}

\label{QWGT}

Quadratically signed weight enumerators (QWGTs) were introduced by Knill and
Laflamme in Ref.~\cite{Knill:01a} (where \textquotedblleft
QWGT\textquotedblright\ is pronounced \textquotedblleft
queue-widget\textquotedblright ). A general QWGT is of the form 
\begin{equation}
S(A,B,x,y)=\sum_{b:Ab=0}(-1)^{bBb}x^{|b|}y^{n-|b|}  \label{eq:QWGT}
\end{equation}%
where $A$ and $B$ are $0,1$-matrices with $B$ of dimension $n\times n$ and $%
A $ of dimension $m\times n$. The variable $b$ in the summand ranges over $%
0,1$-column vectors of dimension $n$, and all calculations involving $A$, $B$
and $b$ are done modulo $2$. It should be noted that $|S(A,B,x,y)| \leq (|x|+|y|)^{n}$.
In Ref.~\cite{Knill:01a} it was shown that
quantum computation is polynomially equivalent to classical probabilistic
computation with an oracle for estimating the value of certain QWGTs with $x$
and $y$ rational numbers. In other words, if these sums could be evaluated,
one could use them to generate the quantum statistics needed to simulate the
desired quantum system.

More specifically, let $I$ be the identity matrix, $\mathrm{diag}(A)$ the diagonal
matrix whose diagonal is the same as that of $A$, and $\mathrm{ltr}(A)$ a matrix formed from the lower triangular
elements of $A$ (the matrix obtained from $A$ by setting to zero all
the elements on or above the diagonal). Then for:

\begin{myproblem}
  \label{KLprob}
\emph{KL promise problem}: Determine the sign of $S(A,\mathrm{ltr}(A),k,l)$
with the restrictions of having $A$ square, $\mathrm{diag}(A)=I$, $k$ and $l$
being positive integers, and the promise $|S(A,\mathrm{ltr}(A),k,l)|\geq
(k^{2}+l^{2})^{n/2}/2$.
\end{myproblem}
KL demonstrated the following:

\begin{mytheorem}
(Corollary 12 in \cite{Knill:01a}): The KL promise problem is BQPP-complete.
\end{mytheorem}

KL's strategy in showing the connection between QWGT evaluation and quantum
computation was to show that in general expectation values of quantum
circuits can be written as QWGTs.

\section{The Partition Function - QWGT Connection}
\label{connection}

We are now ready to prove our central result.

\begin{mytheorem}
\label{th:main}The spin-glass partition function is a special case of QWGTs.
Specifically:%
\begin{equation}
\frac{(1-\lambda ^{2})^{|E|/2}}{2^{|V|}}Z(w)=S(A,\mathrm{dg}(w),\lambda
,1).  \label{eq:same}
\end{equation}
Here $\mathrm{dg}(w)$ is the matrix formed by putting $w$ on the diagonal
and zeroes everywhere else, and $A$ is the incidence matrix of $G$.
\end{mytheorem}

\begin{proof}
In Eq.~(\ref{eq:QWGT}) identify $b$ as the subgraphs of $G=(E,V)$, $n=|V|$, $%
m=|E|$, and note that when $B=\mathrm{dg}(w)$ 
\begin{equation*}
b B b=\sum_{i}b_{i}w_{i}b_{i}=b\cdot w
\end{equation*}%
since $b_{i}=0$ or $1$. Then Eq.~(\ref{eq:same}) follows by inspection of
Eqs.~(\ref{eq:8}) and (\ref{eq:QWGT}).
\end{proof}

\begin{mycorollary}
Evaluating the spin glass partition function is in \#P.\footnote{\#P is the class of function problems of the form ``compute $f(X)$'', where 
$f$ is the number of accepting paths of a nondeterministic polynomial-time
Turing machine. The canonical \#P-complete problem is \#SAT: given a Boolean
formula, compute how many satisfying assignments it has \cite{Aaronson-zoopage,Papadimitriou:book}.}
\end{mycorollary}

\begin{proof}
The problem of evaluating QWGTs at integers is in the class \#P
\cite{Knill:01a}.\footnote{It contains the problem of evaluating the
  weight enumerators of a binary linear code at rational numbers,
  which is \#P-complete \cite{Vertigan:92}.} In our case $x=1$,
$y=\lambda $ and the coupling constant $J$ can always be chosen so
that$\ \lambda =\tanh(\beta J) $ is integer.
\end{proof}

It is tempting to check the relation of Theorem~\ref{th:main} to the
KL promise problem (Problem~\ref{KLprob}). It follows from Eq.~(\ref{eq:same}), from $Z>0$, and from $0\leq
\lambda =\tanh (\beta J)\leq 1$, that $\mathrm{sign}[S(A,\mathrm{dg}%
(w),\lambda ,1)]=+$. Hence, unfortunately, the KL problem in its present
form is of no use to us.

Further consideration reveals that, while the constraint that $k$ and $l$
are positive integers is easily satisfied, and the promise takes a nice
symmetric form:\ $Z(w)\geq 2^{|V|-1}\frac{(1+\lambda ^{2})^{|V|/2}}{%
(1-\lambda ^{2})^{|E|/2}}$, the remaining constraints -- $A$ square, $%
\mathrm{diag}(A)=I$, $B=\mathrm{ltr}(A)$ -- anyhow result in severely
restricted instances of spin glass graphs. We thus leave as open the
following problem, inspired by the KL problem:

\begin{myproblem}
Formulate a promise problem in terms of $Z(w)$ [or, equivalently, 
$S(A,\mathrm{dg}(w),\lambda ,1)$] which is BQPP-complete.
\end{myproblem}

We turn next to showing the connection between our discussion so far and
problems in knot theory.

\section{The Partition Function - Knots Connection}
\label{knots}

The canonical problem of knot theory is to determine whether two given knots
are topologically equivalent. More precisely, in knot theory one seeks to
construct a topological invariant which is independent of the knot shape,
i.e., is invariant with respect to the Reidemeister moves \cite%
{Kauffman:book}. This quest led to the discovery of a number of ``knot
polynomials'' (e.g., the Jones and Kauffman polynomials) \cite{Kauffman:book}%
. These also play a major role in graph theory as instances or relatives of
the dichromatic and Tutte polynomials \cite{Alon:95}, e.g., in the graph
coloring problem. Roughly, two knots are topologically equivalent iff they
have the same knot polynomial. It is well known \cite{Jones:89,Kauffman:book}
that there is a connection between knot polynomials and the partition
function of the Potts spin glass model, a generalization of the Ising spin
glass model to $q\geq 2$ states per spin: 
\begin{equation}
H_{\mathrm{Potts}}(q,{s})=-\sum_{(i,j)\in E}J_{ij}\delta _{s_{i},s_{j}},
\end{equation}
where $s_{i}\in \{0,...,q-1\}$ and $\delta _{s_{i},s_{j}}=1$ ($0$) if $%
s_{i}=s_{j}$ ($s_{i}\neq s_{j}$). We first briefly review this
connection.

\begin{figure}[tbp]
\hspace{10cm} \includegraphics[height=10cm,angle=270]{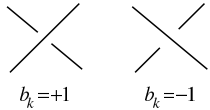} \vspace{%
-6cm}
\caption{Definition of crossing variable for knots.}
\label{fig:crossing}
\end{figure}

Consider a knot embedded in 3D-space (imagine, e.g., a piece of rope). In the
standard treatment \cite{Kauffman:book}, the knot is projected onto the
plane and one obtains a ``2D-knot diagram''. The essential topological
information about the knot is contained in the pattern of ``crossings'', the
2D image of where one rope segment went over or under another rope segment.
A crossing takes values $b_{k}=\pm 1$ according to Fig.~\ref{fig:crossing} 
\cite{Kauffman:87}.

A connection to spin glasses can be made by assigning quenched random values
to the crossing variables $b_{k}$, so that the links cross above and below
at random. It was shown by Nechaev \cite{Nechaev:98} that in this case the
Kauffman polynomial is identical, up to an irrelevant multiplicative factor,
to the Potts model partition function, $Z_{\mathrm{Potts}}(q,\{J_{ij}\})=%
\sum_{s}\exp [-\beta H_{\mathrm{Potts}}(q,{s})]$. To explain this connection
we need to introduce some terminology. The 2D-knot diagram lives on a
lattice $\mathcal{M}$ composed of lines oriented at $\pm 45^{\mathrm{o}}$,
intersecting at the crossings\ $b_{k}$, that carry the disorder. One can
define a dual lattice $\mathcal{L}$, rotated by $45^{\mathrm{o}}$, so that
its horizontal and vertical edges (denoted $b_{ij}$) are in one-to-one
correspondence with the vertices $b_{k}$ of $\mathcal{M}$ (Fig.~6 in \cite%
{Nechaev:98}). Let 
\begin{equation}
b_{ij}=\left\{ 
\begin{array}{c}
-b_{k}\quad \text{\textrm{if the }}(ij)\text{-\textrm{edge is vertical}} \\ 
b_{k}\quad \text{\textrm{if the }}(ij)\text{-\textrm{edge is horizontal}}%
\end{array}%
\right. .
\end{equation}%
The Potts spin states $s_{i}$ are connected to knot properties in an
abstract manner; they are related to the Kauffman polynomial variable $A$,
which in turn is a weight for the manner in which 2D-knot diagram is
disassembled into a set of microstates (and also related to the Jones'
polynomial variable $t$:\ the Jones and Kauffman polynomials coincide when $%
t=A^{1/4}$). Precise definitions can be found in \cite{Kauffman:book}; for
our purposes what matters is that the equivalence of the Kauffman polynomial
to the Potts spin glass partition function is established once one assigns
the Potts variables $q$ and $\beta J_{ij}$ the values 
\begin{eqnarray}
q &=&(A^{2}+A^{-2})^{2}, \notag \\
\beta J_{ij} &=&\ln (-A^{-4b_{ij}}).
\end{eqnarray}%
With these identifications Nechaev has shown that the Kauffman polynomial
(knot invariant) $\langle K(A)\rangle =c(A,\{b_{ij}\})Z_{\mathrm{Potts}%
}(q,J),$ where the constant $c$ does not depend on the spin states \cite%
{Nechaev:98}. Solving for $A$ we find $A=\pm \lbrack (q^{1/2}\pm
(q-4)^{1/2})/2]^{1/2}$. Thus $\beta J_{ij}$ can be real only for $q\geq 4$.
In the Ising spin glass case ($q=2$) we obtain a complex-valued $\beta
J_{ij} $, which in turn implies complex-valued $\lambda =\tanh (\beta J)$,
and hence the estimation of the QWGT polynomial with complex-valued $x$.

Finally, we note that a physically somewhat unsatisfactory aspect of the
knots-Potts connection is that now the (complex-valued) temperature cannot
be tuned independently from the bonds $J_{ij}$. However, this does not matter
from the computational complexity perspective: we have established our third main result: 

\begin{myproposition}
Computing the Kauffman polynomial at $q=2$ is equivalent to the problem of
computing the QWGT polynomial with complex-valued $x$.\footnote{This knots-QWGT connection does not
  appear to hold in the case $q>2$, since in this case we cannot separate $
  \delta _{s_{i},s_{j}}$ into a product of single-spin variables, a step that
  is essential in deriving the representation of $Z$ as a QWGT [see text around
  Eq.~(\protect\ref{eq:2})].} 
\end{myproposition}
Hence an efficient
quantum algorithm for estimating QWGTs will be decisive for knot and graph
theory as well.

\section{Conclusions}
\label{conclusion}
The connection between QWGTs and quantum computational complexity
established by KL on the one hand, and the connection between QWGTs and the
spin glass and knots problems established here on the other hand, suggests
that the quantum computational complexity of spin glass and knots problems
may be decided via the connection to QWGTs. Similar remarks apply to a
number of combinatorial problems in graph theory, via their well-established
connections to knot theory. In particular, it would be desirable to find out
the quantum computational complexity of questions framed in terms of
properties of $S(A,\mathrm{dg}(w),\lambda ,1)$, with $\lambda $ real
(Ising spin glass) or complex (Kauffman polynomial with $q=2$). We leave
these as open problems for future research.

\begin{acknowledgments}
The author thanks the Sloan Foundation for a Research Fellowship and
  Joseph Geraci, Louis Kauffman, Emanuel Knill, Ashwin Nayak, and Umesh
  Vazirani for useful discussions.
\end{acknowledgments}

\appendix

\section{Additional Observations}

\label{further}

\subsection{The case with a Magnetic Field}
\label{appA}

A magnetic field $B$ can be included in the Hamiltonian [Eq.~(\ref{eq:H})],
by adding a term $-\sum_{i\in V}B_{i}\sigma _{i}$. We can repeat the analysis
above by introducing a fictitious \textquotedblleft
always-up\textquotedblright\ spin, numbered $0$. In this manner we can
rewrite the magnetic field term as 
\begin{equation}
\sum_{i\in V}B_{i}\sigma _{i}=\sum_{i\in V}B_{i0}\sigma _{i}\sigma _{0},
\end{equation}
where $\sigma _{0}\equiv 1$. The corresponding graph has a ``star''
geometry, with spin $0$ in the center, connected to all other spins, which
in turn are connected only to spin $0$ (Fig.~\ref{fig:graphs}c). The
analysis above can then be repeated step-by-step, with the relevant
subgraphs being those of the star graph. However, we then cannot recover the
QWGT form, due to the extra summation over the star-graph subgraphs:\ Denote
the latter $b^{\prime }$. Since each spin in $V$ is connected once to the
central spin $\sigma _{0}$, the star graph subgraphs all have trivial parity
vectors, ${\alpha }_{i}^{b^{\prime }}=1$. Then Eq.~(\ref{eq:7}) is replaced
by 
\begin{equation}
\sum_{s}(-1)^{[{\alpha }^{b}+({\alpha }^{b^{\prime }})]\cdot
s}=2^{|V|}\delta _{({\alpha }^{b}+{\alpha }^{b^{\prime }},0)}.
\end{equation}
This causes a violation of the condition $b:Ab=0$ needed in the definition
of the QWGT sum. Thus it appears that QWGTs do not include the case with a
magnetic field.

\subsection{Power Series Representation}
\label{appB}

Another useful representation of Eq.~(\ref{eq:8}) can be obtained by
grouping together all subgraphs with the same number of edges. To this end,
let $b_{j}^{(k)}$ denote the $j^{\mathrm{th}}$ subgraph with $k$ edges.
According to the numbering scheme introduced in Sec.~\ref{Fourier}, the corresponding binary
number of such a subgraph, $b_{j}^{(k)}$, is the $j^{\mathrm{th}}$
permutation of a vector of exactly $k$ 1's and $|E|-k$ zeroes. Since $|b|$
is the Hamming weight of $b$ these subgraphs all have $|b| = k$. There are $%
\binom{|E|}{k}$ such subgraphs, all with equal weight $\lambda ^{k}$.
Therefore:
\begin{eqnarray}
&&  Z(w)={\frac{2^{|V|}}{(1-\lambda ^{2})^{|E|/2}}} \times \nonumber \\
&&  \left( 1+\sum_{k=1}^{|E|}\lambda ^{k} \sum_{j=1}^{ \binom{|E|}{k}%
}(-1)^{b_{j}^{(k)}\cdot b} \delta _{({\alpha }^{b_{j}^{(k)}},0)} \right)
.  \label{eq:9}
\end{eqnarray}
In this form we have a series expansion in powers of $\lambda $,
corresponding to the number of edges of the subgraphs.

A clear simplification results in the fully ferromagnetic Ising model ($%
Z_{+} $), where $w\equiv (0,0,...,0)$, and in the fully antiferromagnetic
case ($Z_{-}$), where $w\equiv (1,1,..,1)$. In the latter case we have simply $%
b\cdot w = |b|$, so that combining the two cases we obtain from Eq.~(\ref{eq:8}):

\begin{equation}
Z_{\pm }={\frac{2^{|V|}}{(1-\lambda ^{2})^{|E|/2}}}\sum_{b}(\pm
\lambda)^{|b|}\delta _{({\alpha }^{b},0)}.
\label{eq:F-AF}
\end{equation}
Eq.~(\ref{eq:9}), on the other hand yields:

\begin{equation}
Z_{\pm }={\frac{2^{|V|}}{(1-\lambda ^{2})^{|E|/2}}}\left(
1+\sum_{k=1}^{|E|}(\pm \lambda )^{k}\sum_{j=1}^{\binom{|E|}{k}}\delta _{({\
    \alpha }^{b_{j}^{(k)}},0)}\right) .
\label{eq:10}
\end{equation}
As already remarked, there exist an efficient classical algorithm for
calculating $Z$ in the case of the fully ferromagnetic Ising model \cite{Jerrum:90}.


\end{document}